\documentclass{article}
\usepackage{spconf,amsmath,graphicx,subfigure}
\usepackage{tabularx}

\title{Reply to Cox et al. and Kessler et al.: Data and code sharing is the way forward for fMRI}

\name{Anders Eklund$^{\: a,b,c}$, Thomas Nichols$^{\: d}$, Hans Knutsson$^{\: a,c}$}
\address{$^a$Division of Medical Informatics, Department of Biomedical Engineering, \\ Link\"{o}ping University, Link\"{o}ping, Sweden \\ $^b$Division of Statistics and Machine Learning, Department of Computer and Information Science, \\ Link\"{o}ping University, Link\"{o}ping, Sweden \\ $^c$Center for Medical Image Science and Visualization (CMIV), \\ Link\"{o}ping University, Link\"{o}ping, Sweden \\ $^d$Department of Statistics \& WMG, University of Warwick, Coventry, United Kingdom}

\begin{document}
%
\maketitle

We are glad that our paper~\cite{eklund} has generated intense discussions in the fMRI field, on how to analyze fMRI data and how to correct for multiple comparisons. The goal of the paper was not to disparage any specific fMRI software, but to point out that parametric statistical methods are based on a number of assumptions that are not always valid for fMRI data, and that non-parametric statistical methods~\cite{winkler} are a good alternative. Through AFNI's introduction of nonparametric statistics in the function 3dttest++~\cite{cox,cox2}, the three most common fMRI softwares now all support non-parametric group inference (SPM through the toolbox SnPM (http://warwick.ac.uk/snpm), and FSL through the function randomise).

\section{Reply to Cox et al.}

Cox et al.~\cite{cox} correctly point out that the bug in the AFNI function 3dClustSim only had a minor impact on the false positive rate (FPR). This was also covered in the original paper~\cite{eklund}, "We note that FWE rates are lower with the bug-fixed 3dClustSim function. As an example, the updated function reduces the degree of false positives from 31.0\% to 27.1\% for a CDT of P = 0.01, and from 11.5\% to 8.6\% for a CDT of P = 0.001." It is unfortunate that several media outlets focused extensively on this bug, when the main problem was found to be violations of the assumptions in the statistical models. 

The statement that AFNI had particularly high FPRs, compared to SPM and FSL, is for example supported by Supplementary Figure 1 a~\cite{eklund} (Beijing data, two-sample t-test with 20 subjects, cluster defining threshold p = 0.01). For 8 mm of smoothing, the FPR for AFNI is 23\% - 31\%, while it is 13\% - 20\% for SPM and 14\% - 18\% for FSL OLS. To understand the higher FPRs we investigated how the 3dClustSim function works, which eventually lead us to finding the bug in 3dClustSim. However, we agree that AFNI did not produce higher FPRs for all parameter combinations.

The 70\% FPR comes from Supplementary Figure 9 c~\cite{eklund} (Oulu data, one-sample t-test with 40 subjects, cluster defining threshold p = 0.01, FSL OLS with 4 mm smoothing), and not, as some readers believed, from Figure 2 in the original paper~\cite{eklund} which shows results for the ad-hoc clustering approach. The main reason for using the highest observed FPR was to give the reader an idea of how severe the problem can be, but we agree that it lead to a too pessimistic view.

As pointed out by Cox et al.~\cite{cox}, the non-parametric approach also performed sub-optimal for the one-sample t-test, especially for the Oulu data. As discussed in our paper, the one-sample t-test has an assumption of symmetrically distributed errors that can be violated by outliers in small samples. Our current research is therefore focused on how to improve the non-parametric test for one-sample t-tests. Regarding the flexibility of the permutation testing, recent work has shown that virtually any regression model with independent errors can be accomodated~\cite{winkler}, and even longitudinal and repeated measures data can be analyzed with a related bootstrap approach~\cite{guillaume}.

\section{Reply to Kessler et al.}

Kessler et al.~\cite{kessler} extend our evaluations to (non-parametric) cluster based false discovery rate (FDR) on task data, to better understand how existing parametric cluster p-values based on the familywise error rate (FWE) should be interpreted. For the problematic cluster defining threshold of p = 0.01, they conclude that a cluster FWE-corrected p-value smaller than p = 0.00001 survives FDR correction at q = 0.05. Indeed, this information makes it easier to interpret existing results in the fMRI literature, but it should be noted that it is not straightforward to generalize these results to other studies. For example, the fMRI software used, the MR sequence used (EPI or multiband), the degree of smoothing and the number of subjects are all likely to affect this cutoff. The only way to retrospectively evaluate existing results is, in our opinion, to re-analyze the original fMRI data (e.g. made available through OpenfMRI~\cite{openfmri}) or to apply a new threshold to the statistical maps (e.g. made available through NeuroVault~\cite{neurovault}).

\section{Importance of data and code sharing}

Cox et al.~\cite{cox,cox2} replicated and extended our findings with the same open fMRI data~\cite{biswal2} as in our original paper (and made use of our processing scripts available on github, https://github.com/wanderine/ParametricMultisubjectfMRI), ultimately resulting in improvements to the AFNI software. Further, we never would have been able to identify the bug in 3dClustSim were AFNI not open source software. Kessler et al.~\cite{kessler} also used the same task datasets from OpenfMRI~\cite{openfmri}, to find the empirical cluster FDR. Together, these examples show the importance of data sharing~\cite{poldrack,poline}, open source software~\cite{ince}, code sharing~\cite{baker,eglen} and reproducibility~\cite{bids}.

\section*{Acknowledgements}

This research was supported by the Neuroeconomic Research Initiative at Link\"{o}ping University, by Swedish Research Council Grant 2013-5229 ("Statistical Analysis of fMRI Data"), the Information Technology for European Advancement 3 Project BENEFIT (better effectiveness and efficiency by measuring and modelling of interventional therapy), the Swedish Research Council Linnaeus Center CADICS (control, autonomy, and decision-making in complex systems), and the Wellcome Trust.

\bibliographystyle{IEEEbib}
\bibliography{references}

\end{document}